\documentclass[a4paper,fleqn,usenatbib]{mnras}

\usepackage{txfonts}

\usepackage[T1]{fontenc}
\usepackage{ae,aecompl}

\usepackage{graphicx}	
\usepackage{amssymb}	
\usepackage{wasysym}

\title{ZAP -- Enhanced PCA Sky Subtraction for Integral Field Spectroscopy}

\author[K. T. Soto et al.]{
Kurt T. Soto$^{1}$\thanks{E-mail: kurt.soto@phys.ethz.ch},
Simon J. Lilly$^{1}$,
Roland Bacon$^{2}$,
Johan Richard$^{2}$,
Simon Conseil$^{2}$
\\
$^{1}$ETH Zurich, Institute of Astronomy, Wolfgang-Pauli-Str. 27, CH-8093 Zurich, Switzerland\\
$^{2}$Univ Lyon, Univ Lyon1, Ens de Lyon, CNRS, Centre de Recherche Astrophysique de Lyon UMR5574, F-69230, Saint-Genis-Laval, France \\
}

\date{Accepted 2016 February 25. Received 2016 February 25; in original form 2016 January 15}

\pubyear{2016}

\begin{document}
\label{firstpage}
\pagerange{\pageref{firstpage}--\pageref{lastpage}}
\maketitle

\begin{abstract}
We introduce Zurich Atmosphere Purge (ZAP), an approach to sky subtraction based on principal component analysis (PCA) that we have developed for the Multi Unit Spectrographic Explorer (MUSE) integral field spectrograph. ZAP employs filtering and data segmentation to enhance the inherent capabilities of PCA for sky subtraction. Extensive testing shows that ZAP reduces sky emission residuals while robustly preserving the flux and line shapes of astronomical sources. The method works in a variety of observational situations from sparse fields with a low density of sources to filled fields in which the target source fills the field of view. With the inclusion of both of these situations the method is generally applicable to many different science cases and should also be useful for other instrumentation. ZAP is available for download at \mbox{\texttt{http://muse-vlt.eu/science/tools}}. 
\end{abstract}



\section{Introduction}

A number of techniques have been developed to address the problem of sky subtraction for ground-based spectroscopic measurements of faint objects. In most cases the methods leave behind narrow residual features which can hinder the detection and measurement of astronomical sources. These residuals come from the oversubtraction and undersubtraction of a modeled or interpolated sky emission spectrum. The residuals arise due to small variations in the characteristics of the instrument at different spatial positions and thus associated imperfections in the calibration process.  Sky residuals are particularly problematic where the background is rapidly changing with wavelength, i.e. around sky emission lines, and when the spectrum is under-sampled by the detector.
 
For long-slit spectrographs, the sky is often removed using a spatial interpolation between blank sky regions in the slit adjacent to sources. While this often works well, it is limited to sources that have a smaller angular size than the extent of the long-slit. Although it would appear to be directly applicable to the data cube produced by a large format two dimensional integral field spectrograph (IFS), in practice the elaborate mapping between the detectors and the reconstructed datacubes introduces small discontinuities in the line spread function (LSF), wavelength solution, and flat fielding. These small discontinuities make such spatial interpolation problematic.

Fibre spectrographs are subject to similar problems. Each individual fibre will have a different LSF due to the different fibre and light paths.  However, post processing with principal component analysis (PCA) has been effective at removing residuals in the spectra from large fibre based surveys \citep{kurtz2000,wild2005,sharp2010}.

The MUSE integral field spectrograph (IFS) \citep{MUSE} uses an image slicing optical component called a ``field splitter'' to separate the field of view into 24 regions that are passed into 24 identical but independent integral field units (IFUs). Within each of these IFUs, the separate fields of view are again split by an image slicer into 48 slit-like image slices. Light in these slices passes through a volume phase holographic grating and is dispersed onto the detector.  Ultimately the combination of slices and IFUs are reconstructed into a complete datacube of 90,000 spectra, with a $1' \times 1'$ field of view sampled at $0.2''$ per spaxel and covering an octave of wavelength from 4750 \AA\ -- 9350 \AA\ sampled at 1.25 \AA\ per pixel.

The multiple IFU, image slicer design of the MUSE spectrograph, while providing the benefit of a large contiguous field of view and an octave of spectral coverage with essentially no moving parts, introduces a few unique challenges in performing sky subtraction. The major difficulty in performing this task is that the light path varies from slice to slice and from IFU to IFU. This variation in light path  introduces small discontinuous variations in the LSF and the wavelength solution of the final reconstructed datacube over relatively small spatial regions. As a result, most general, and even some localized sky subtraction methods may mischaracterize the sky signal in the spaxels that contain astronomical signals.

The standard MUSE data reduction pipeline \citep{weilbacher2012,weilbacher2014} includes an advanced sky-subtraction algorithm that uses a physical model of the sky emission to remove the sky signal \citep{streicher2011}.
This approach, however, requires very precise knowledge of the LSF and accurate wavelength solution, both of which are challenging to achieve given the nonuniform LSF sampling and variation of the LSF in a single MUSE exposure. This sampling can range from 0.3 to 0.55 \AA\ on the detector itself, while the final reconstructed datacube is sampled at 1.25 \AA, which translates to a less than Nyquist sampling of the 2.3 \AA\ LSF in the final datacube. 
Furthermore, small variations in the flat field and bias level can cause small systematic variations relative to the average sky.  
The consequence is that the current sky subtraction method leaves behind significant residuals that can interfere with faint source detection, especially in the spectral regions populated by strong sky emission lines.

There is therefore interest in developing empirical post-processing methods that do not require a precise knowledge of the instrument LSF and that can cope with the inaccuracies introduced by flat field residuals. Among the possible methods, principal component analysis (PCA) is an obvious choice given its ability to characterize the correlated variations from a large sample of measurements.  An important point is that the sky residuals are likely to be correlated over a small range of wavelength within a given spectrum.  This is because the imperfections that cause them, e.g. imperfect wavelength calibration and LSF construction or even just the under-sampled mapping of the detector onto the spectrum, should vary slowly with wavelength along a given spectrum.  A PCA approach can therefore use the very large number of (normalized) spectra from different spaxels within the datacube to calculate the eigenspectra of the residual sky features.

A risk in using standard PCA, however, is that the eigenspectra constructed to characterize the residuals are unable to distinguish between astronomical signals (such as emission lines in spectra) and the sky. Without careful consideration of this potential problem, a PCA approach can corrupt the flux of the astronomical sources. Several of the details of the implementation presented here are designed to minimize and test this issue.

To meet the challenge of reducing sky residuals from the 90,000 observed spectra in each single exposure with the MUSE spectrograph, we have developed the Zurich Atmosphere Purge (ZAP) algorithm. 
ZAP is based on PCA and is designed to minimize the effects on astronomical source flux. The preservation of astronomical source flux is achieved through a combination of spectral filtering and data segmentation algorithms at particular points in the process. By including these steps we minimize or eliminate the astronomical signals from the set of spectra that are used to calculate the eigenbasis, which therefore creates a ``sterilized'' set of eigenspectra that only consists of sky emission features.

The ZAP algorithm was primarily developed for the case of small faint sources in the MUSE field of view, the ``sparse-field'' case below. However, with some modification it also works very well in the case where an object of large angular size fills the field of view, provided that a ``blank-field'' datacube is available. We will refer to this case as the ``filled-field" case.

This paper describes the philosophy, implementation, and testing of these methods and ends with a discussion of the limitations of the ZAP sky subtraction method. Although developed for the particular case of MUSE, the filtering and data segmentation algorithms should be generally applicable to any dataset with similar residuals including the single fibre spectra from existing surveys with the  Anglo-Australian Telescope (AAT) \citep{sharp2010}, Sloan Digital Sky Survey (SDSS) \citep{Smee2013}, Calar Alto Legacy Integral Field Area Survey (CALIFA)\citep{CALIFA}, and with current and future instruments such as the Mapping nearby Galaxies at Apache Point Observatory survey (MaNGA) \citep{MANGA}, the Prime Focus Spectrograph (PFS) \citep{PFS}, the Multi-Object Optical and Near-infrared Spectrograph (MOONS)\citep{MOONS} or the 4-metre Multi-Object Spectroscopic Telescope (4-MOST)\citep{4MOST}.  In fact, because ZAP does not use the relative spatial information in the cube (except for the masking of known sources) ZAP could be used for single fibre spectral surveys without much modification.

\section{Method}
\label{sect:Method}

\subsection{Overview}
\label{sect:overview}
\begin{figure*}
\centering
\includegraphics[width=.9\linewidth]{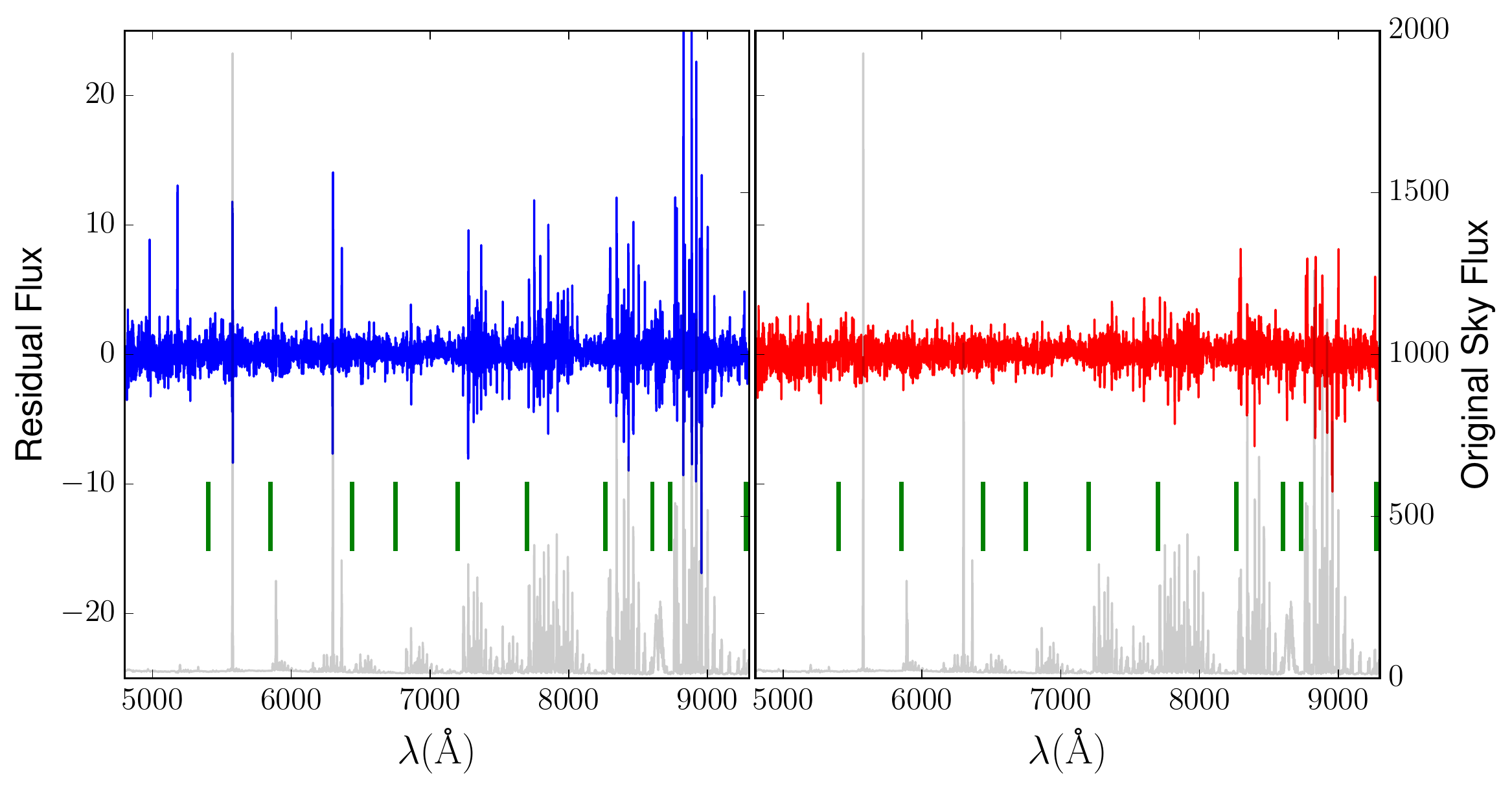}
\caption{\label{fig:skyspec} The mean flux over a circular aperture of 1 arcsec diameter located in an empty region of a 25 min single exposure of the Ultra Deep Field (UDF). The fluxes are all in $10^{-20}~$erg s$^{-1}$cm$^{-2}$, with the scale for the remaining sky residuals on the left vertical axis and the scale for the original sky spectrum (plotted in gray) on the right vertical axis. The residuals left behind after the standard sky subtraction routine can exceed the Poisson error by factors of 1.2 to 2.0 at the wavelengths of strong sky emission lines, while ZAP is able to keep the spectrum within the Poisson error throughout the spectrum.
In both panels, the green line shows the separation into spectral segments described in Section \ref{sect:segmentation}. {\it Left:} Standard Processing (specifically version 1.0 of the pipeline) {\it Right:} ZAP.}
\end{figure*}

As described above, ZAP is designed to subtract residual sky features while minimizing the effect on astronomical signals. 
Through the use of principal component analysis combined with filtering, ZAP constructs a sky residual spectrum for each individual spaxel which can then be subtracted from the original data cube.

ZAP operates by first applying a coarse sky subtraction either to a cube produced by a preliminary sky-subtraction, or in fact from a cube that has not had the sky subtracted.  This leaves behind residuals which are the result of variations in the LSF, wavelength solution, and flatfielding over the field.  
Astronomical signals in this cube are then reduced (or even eliminated) by wavelength filtering and/or spatial masking, leaving behind a relatively clean set of input spectra that are dominated by sky residual features.
Once this clean set of spectra is established, ZAP performs singular value decomposition (SVD) simultaneously in a set of discrete spectral segments.
In the nominal MUSE octave from 4750 \AA\ to 9350 \AA\ we use 11 spectral segments that together span the whole spectral range.
Using these eigenspectra, ZAP then calculates the eigenvalues for each individual spectrum, having applied a second weighted median filter to further reduce any astronomical signal in the spectrum.  Finally, ZAP calculates the overall variance (within the subcube associated with a spectral segment) that is associated with each successive eigenspectrum and autonomously chooses the optimal number of eigenspectra to use in the reconstruction of the residual spectra. 
In order to improve the spatial mask used to remove detected sources, ZAP can also be applied iteratively to further reduce the residual sky features.

We detail these steps in the following sections.
Using post-processing by ZAP, we are able to reduce the sky emission residuals that are present in the data cubes produced by the MUSE pipeline as illustrated in Figure \ref{fig:skyspec}.

\subsection{Prefiltering}
\label{sect:prefiltering}

One problem with the blind application of SVD is that the orthonormal basis created by this process can entangle astronomical source signals and sky emission signals.
To circumvent this entanglement, we apply a series of filters and further segment the data in wavelength to isolate the sky signal in the constructed eigenspectra.  In the following section we describe these preparatory steps which sterilize the dataset that is used to construct the eigenspectra.

\subsubsection{Masks}
\label{sect:masks}
	
The simplest approach to eliminate the contribution of astronomical sources to the constructed eigenspectra is to spatially mask them from the start by removing these spaxels. 
The necessity of this step depends on the spatial coverage of the astronomical source(s) as well as their flux relative to the sky.
In the case of very faint sources, this step is found to be not strictly necessary; however, masks can be employed to improve the result and to guarantee that these spaxels are not contribute to either the zeroth order sky spectrum, described in Section \ref{sect:zorder}, or to the eigenspectra, described in Section \ref{sect:eigenvectors}. 
A simple masking of sources that appear in white light images will often be sufficient, but the user can define any spatial mask to ensure that these spaxels do not contribute. In most cases masking of objects that are 2$\sigma$ above the background will be sufficient. In the case of science that is focused on extended emission in spectral lines, it is helpful to include strong detections at these wavelengths to further ensure their exclusion from the zeroth order sky spectrum and the eigenspectra.

When ZAP is used iteratively, improvement of the masks becomes crucial to the overall improvement in performance. 
As the sky residuals are removed, the contribution of any unmasked sources will increase and they may eventually become the dominant contributors to the variance and thus dominate the eigenspectra.
By masking the sources more aggressively, this problem can be circumvented, and the sky emission residuals can be further improved.

\subsubsection{Zeroth order sky spectrum}
\label{sect:zorder}

The zeroth order sky subtraction step removes the median sky signal as a coarse first step in sky subtraction.
This zeroth order spectrum is found by calculating the median (or optionally an iteratively sigma-clipped mean) for each monochromatic layer of the input datacube.
This approach removes 99\% of the overall sky signal but leaves behind large residuals in each spaxel that are, as discussed earlier, generally associated with small variations in the line spread function and the wavelength solution over the whole field. This first step can also be applied to a cube that has gone through the standard sky-subtraction pipeline.
By performing this zeroth order sky subtraction the residual sky features become the deviations from the mean that will be characterized by the eigenspectra produced by the SVD. 

	\subsubsection{Source/Continuum Filtering}
	\label{sect:cfilter}

In an effort to further minimize the influence of any astronomical signals on the eigenspectra, we next subtract from each spectrum a filtered version of itself. This step is designed to keep the narrow sky residuals, but to reduce or even eliminate the continuum and those broader spectral features in the astronomical object spectra.  For a moderate resolution spectrograph like MUSE, astronomical signals are usually substantially broader than the sky residuals.
We found that the filter that best handles this requirement is a weighted median filter.

The weighted median filter operates on each spectrum and is designed to de-weight those pixels associated with sky emission lines. 
Using a weighted median allows the filter to trace discontinuities in spectral shape, such as continuum breaks while acting to preserve sky residuals. 
We create the filter by assigning a positive inverse weight to each pixel associated with the flux of the zeroth order sky spectrum at the same wavelength, using the absolute value in the case that a pre-sky-subtracted cube is used. 
We then sort the pixels within the filter window according to the flux.  The weighted median is then found in the usual way by identifying the flux value that is coincident with the midpoint of the cumulative sum curve of the weights.  
In the case of equal weights this value is the same as the standard median. 

\subsection{Principal Component Analysis}
\label{sect:pca}

The central concept of PCA is to use the data itself to find the eigenvectors (i.e. in this case the eigenspectra) that optimally describe the variations over the entire dataset, ranked in their contribution to the overall variance.
Having sterilized the input dataset (Sect. \ref{sect:prefiltering}), we perform SVD to construct a ranked orthonormal basis of eigenvectors that will characterize the input dataset. 
The full set of eigenspectra will completely define the entire input dataset, including any remaining astronomical signals and even noise.  However, the sterilization process described above should have resulted in the strongest features remaining in the input cube being associated with the sky residuals which will therefore be described in the first few eigenspectra.  In the following sections we detail the application of these steps as used to address sky subtraction with ZAP. 

\subsubsection{Spectrum Segmentation}
\label{sect:segmentation}
		
An important feature of the residuals is that they tend to be correlated in wavelength, over at least a short interval of wavelength. 
This is because the imperfections in the data processing, due to very small inaccuracies in wavelength solution, flat field, and the effects of variation of the LSF are likely to be coherent over a short wavelength interval.  A further reason for correlation is that many of the sky emission lines are produced by families of transitions between the excitation states of OH and, to a lesser extent, O$_2$, O I, and Na I in the atmosphere \citep{osterbrock1996,loo2007}.  These produce the well-known appearance of a forest of sky emission lines which vary coherently in time, separated by regions of much lower sky emission. The correlation in the intensity of sky emission lines associated with vibrational bands has also been observed and used to aid in sky subtraction for spectroscopy in the near infrared \citep{Davies2007}.

We take advantage of this correlation by segmenting the datacube into wavelength sections as listed in Table \ref{table:bins}.
This segmentation has three important advantages. 

First, by restricting attention to a set of residuals that are expected to be varying coherently, the information on the residuals is concentrated into a small number of eigenspectra.  Treating the full wavelength range, which will contain a lot of uncorrelated residual features, will tend to diffuse the information about a given residual feature over a larger number of eigenspectra.

Second, for practical computation purposes, the different segments can be operated on simultaneously through the use of parallel processing. 
This reduces the computation time by more than a factor of the number of bins, since the compute time for SVD scales quadratically with the size of the smallest axis.

Finally, from a data quality standpoint, the segmentation also alters the dimensionality of the input datacubes for each SVD calculation. 
This segmentation reduces the size of the spectral axis in pixels which, as a consequence of linear algebra, is equal to the number of eigenspectra calculated in the segment.
With segments of $\sim 500$ spectral pixels and 90,000 spectra (one per spaxel), the number of eigenspectra calculated will generally be two orders of magnitude smaller than the number of spectra, compared to only 1 order of magnitude for the 3,680 pixel full spectrum.
This fact ensures that there are many measurements of the same correlated residuals, which ensures that they have the most power and are in the first few high ranked eigenspectra.

To further justify the above points about computation time and data quality, it is worth looking at the distribution of ``information'' into the segments via the dimensionality of the data and matrices produced by SVD. Since ZAP does not operate using spatial coherence, the datacubes can be collapsed into a single large $n \times m$ matrix with n spectral pixels ($n=3,680$) and m spectra ($m=90,000$). If the eigenspectra were calculated for this entire wavelength range at once, the result would be an $n \times n$ eigenspectrum matrix with $n \times m$ associated eigenvalues. Under this calculation, the computation time scales as $mn^2$.  In the segmented version of this calculation, the spectral dimension $n$ is reduced by a factor of the number of segments (in our case 11), which we will call $n_i$. The total number of eigenspectra for each segment is then $\sum_{i}{n_i} = n$, which is the same number of eigenspectra. The number of spectra per segment is the same value as the full matrix calculation $m$. The total number of eigenvalues then is $\sum_{i}{n_i} \times m = n \times m$, which shows that there is no loss of information. The benefit however is that each of the segments can be calculated independently, so the calculation time scales as $mn_i^2$, which in our case is equivalent to a factor of $\sim 100$ improvement in calculation time.

\begin{table}
\caption{Wavelength Segments \label{table:bins}}             
\centering                          
\begin{tabular}{l l l l}        
\hline\hline                 
segment &        
start $\lambda$ (\AA~) &        
end $\lambda$ (\AA~) &        
origin \\        
\hline                        
1   & 4750   & 5400   & [OI]  \\
2   & 5400   & 5850   & OH 7-1  \\
3   & 5850   & 6440   & Na I, OH 9-3, [OI]  \\
4   & 6440   & 6750   & OH 6-1  \\
5   & 6750   & 7200   & OH 7-2 \\
6   & 7200   & 7700   & OH 8-3, OH 4-0  \\
7   & 7700   & 8265   & OH 9-4, OH5 5-1  \\
8   & 8265   & 8602   & OH 6-2  \\
9   & 8602   & 8731   & O$_2$  \\
10 & 8731   & 9275   & OH 7-3  \\
11 & 9275   & 9350   & OH 3-0  \\
\end{tabular}
\end{table}

		\subsubsection{Eigenvector Calculation}
		\label{sect:eigenvectors}

\begin{figure*}
\centering
\includegraphics[width=0.85\linewidth, keepaspectratio=true, trim={0 15 0 15},clip]{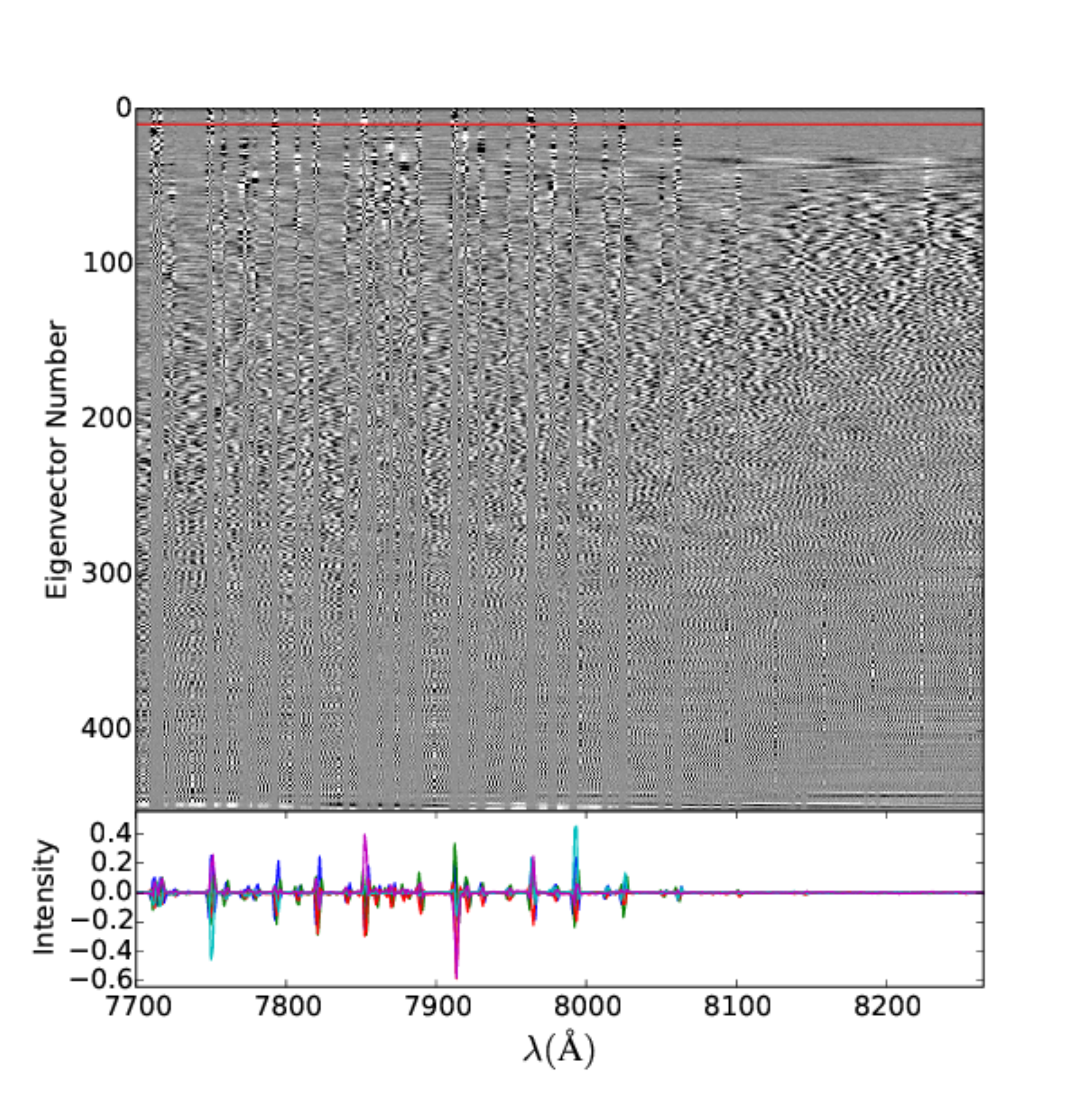}
\caption{\label{fig:evecmatrix} {\it Above:} Matrix of eigenspectra for one particular segment. The vertical axis represents the rank of the eigenvector, with the most significant at the top, and the least significant at the bottom. The number of calculated eigenspectra is determined by the number of pixels in the wavelength direction. The apparent discrepancy between the number of eigenspectra produced and the range on the wavelength axis is due to the 1.25\AA\ per pixel sampling. The red line shows the typical number of spectra used to reconstruct the sky residual spectrum in this segment. 
 {\it Below:} The first 10 eigenspectra plotted together.
 }
\end{figure*}

When we calculate the set of eigenspectra for each segment, we must first normalize the input spectra. For ZAP we chose to normalize each spectrum by its own variance within that segment.

For each segment, we therefore deconstruct the input datacube 
of m spectra, each of n spectral pixels, into an $n \times m$ matrix $A$.
In the case of MUSE we will have $m \gg n$, especially with the segmentation in wavelength as described above. The input matrix $A$ is rectangular, so the identification of the eigenspectra must be performed using singular value decomposition (SVD). The SVD of the matrix $A$ constructs the matrices $U, s, V$ such that the equation 
\[A = U\ s\ V^T\]
holds. In this equation, $U$ is an $m \times m$ matrix where the columns are analogous to the eigenvectors along the spatial axis. The matrix $V^T$ is an $n \times n$ matrix composed of the eigenvectors along the spectral axis. The matrix $s$ is a diagonal matrix of size $n \times m$. This diagonal matrix is composed of the singular values, which are similar to eigenvalues, but require both bases $U$ and $V$ to reconstruct the input matrix $A$.

When these matrices are calculated, the $s$ matrix will be diagonal up to the shape $n \times n$, and beyond that the matrix is entirely zero, which means that the last $m-n$ vectors in the $U$ matrix are unnecessary to reconstruct $A$.  This fact allows us to use the ``reduced'' form of SVD, which leaves out the calculation of these vectors and a therefore accelerates the calculation. Furthermore, since this algorithm is oriented toward finding the eigenvectors along the spectral direction, the $U$ and $s$ matrices are unnecessary for the final product of ZAP. We find the desired eigenvectors, which we call eigenspectra, along the rows of V$^T$\citep{cuppen1983,golub1996,NRC}.

\begin{figure}
\centering
\includegraphics[width=\linewidth, keepaspectratio=true, trim={30 10 50 35},clip]{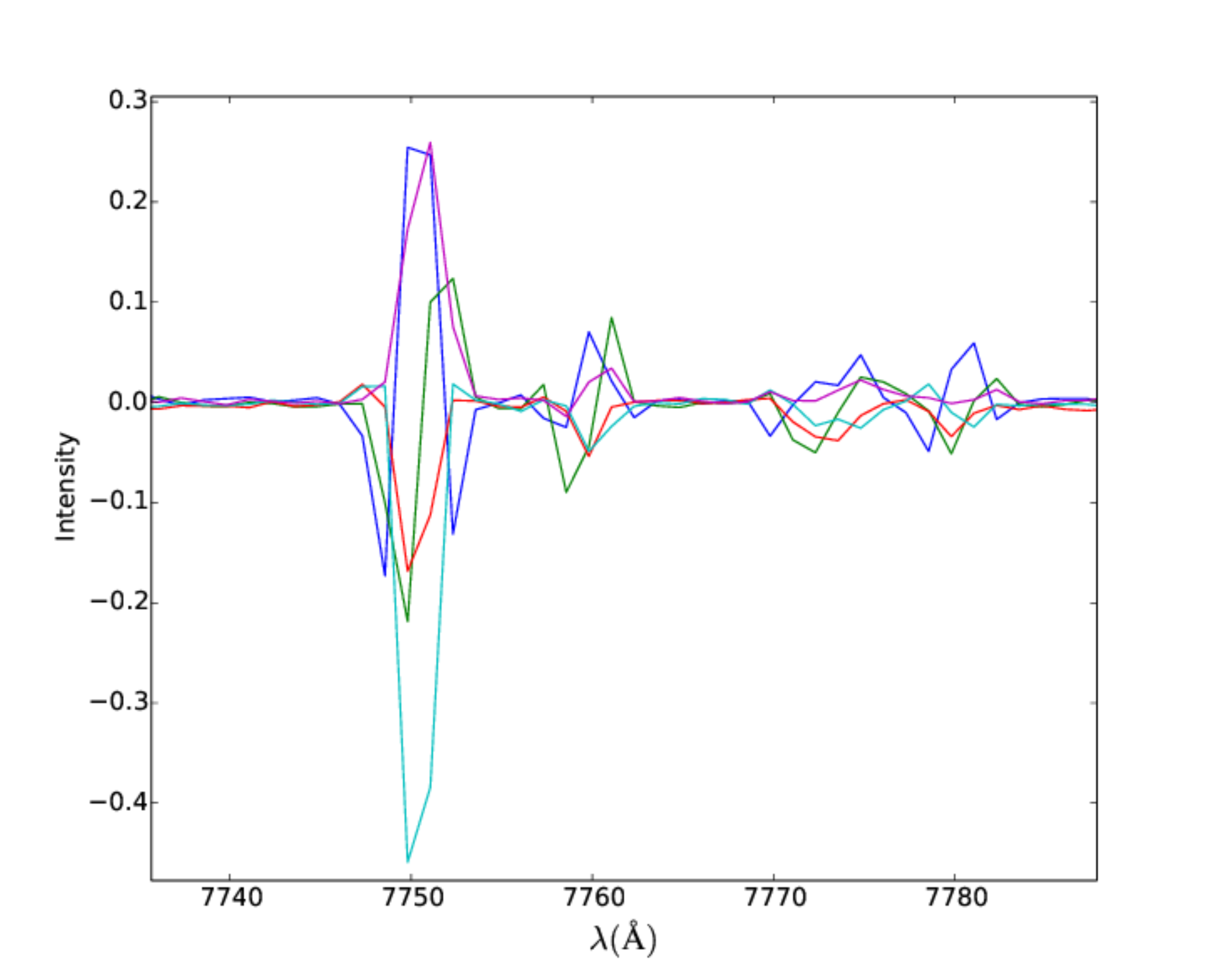}
\caption{\label{fig:zoomespec} The first few eigenspectra that represent sky subtraction residuals at the wavelength of an emission line. The correlation between the residuals at different wavelengths allows the subtraction to operate when source flux is coincident with a sky residual feature, while simultaneously preserving the flux in an astronomical object.}
\end{figure}

The SVD calculation is performed using standard routines in the \texttt{numpy.linalg} library.
We illustrate the output of this calculation in Figures \ref{fig:evecmatrix} and \ref{fig:zoomespec}. In Figure \ref{fig:evecmatrix} we show the matrix of eigenvectors produced from the calculation for one particular segment. 
In the greyscale image each row represents one of the calculated eigenspectra. 
The x axis therefore represents wavelength and the y axis represents the rank of a particular eigenspectrum. 
The red line shows the typical number of eigenspectra used in the construction of the sky residual spectra (see section \ref{sect:optimization}) in the sense that only eigenspectra above the red line are used.  
In the lower panel we plot these 10 eigenspectra together.   The continuum level between the sky residual features in each of these eigenspectra is very close to 0.  When projected onto the data, these spectra will therefore characterize the residuals at specific wavelengths and not affect emission in the regions where the eigenspectra are 0.

In Figure \ref{fig:zoomespec}, we show a close up of a small region of wavelength space, where the eigenspectra characterize a particular group of sky residuals. 
In each eigenspectrum the emission at different wavelengths is linked, such that a correction at 7750 \AA\ will affect the residuals at 7760 \AA\ and 7770 -- 7780 \AA. Each of the eigenspectra characterize a slightly different aspect of where the initial sky subtraction was imperfect. For example, in the blue eigenspectrum there is a central peak with troughs around each of the lines. This eigenspectrum will correct an undersubtraction in the peak of the line and an oversubtraction in the wings of the line. For the green spectrum, the shape has a peak at slightly longer wavelengths and a trough at slightly shorter wavelengths. This eigenspectrum will characterize a slight offset in wavelength solution.  The various eigenspectra put together will characterize all of the aspects of the LSF that were not corrected in the initial subtraction.

\subsubsection{Eigenvalue Calculation}
\label{sect:eigenvalues}
			
In an SVD calculation, the eigenvalues are produced along with the eigenspectra. However, when strong source emission is coincident with sky emission residuals, it is possible that the eigenspectra will be misapplied. This aspect is particularly troublesome in the ``filled-field'' case that is described in Section \ref{sect:filledfield} below.

The eigenvalues for the different eigenspectra that are then used to reconstruct the residual sky spectra in the datacube can in principle then be calculated by taking the dot product of each spectrum (including any previously masked spaxels) with each eigenspectrum.   However, as with the cleaning process above, it is found to be advantageous if the spectra are first filtered with another weighted-median filter before the eigenvalue is calculated.  This can be chosen to be narrower than the filter used to sanitise the spectra before the construction of the eigenspectra, further ensuring that astronomical features are not included in the eigenvalue calculation. This is discussed further in Section \ref{sect:cfwidth} and Section \ref{sect:discussion}.

\subsection{Optimization of the number of eigenspectra to be used}
\label{sect:optimization}

In PCA, one of the difficulties is determining the number of eigenvectors to use. In this case, this corresponds to the number of eigenspectra that will be needed to construct the residual sky spectrum (for each individual spaxel) that will then be subtracted from the original input cube.
Using too many eigenspectra would lead to removal of astronomical signals and to spurious de-noising of the data.  Of course, use of all of the eigenspectra would remove everything in the cube that had not been earlier removed by the pre-filtering.  For ZAP, we have developed an autonomous algorithm to select the number of eigenspectra to be used in the construction of the ``residual sky spectrum''.

The algorithm operates by first producing overall variance curves for each of the spectral segments. These variance curves are created by calculating the total variance across all spectra within the segment in question. We successively increase the number of eigenspectra used to construct the residual sky spectra and examine the consequent reduction in variance in the cube once this is subtracted. This curve decreases monotonically with the number of eigenspectra. Generally, there is a sharp drop after the first few eigenspectra followed by a shallow linear slope as the number of eigenspectra used approaches the total number of eigenspectra. 

\begin{figure}
\centering
\includegraphics[width=\linewidth, keepaspectratio=true,trim={20 45 20 60},clip]{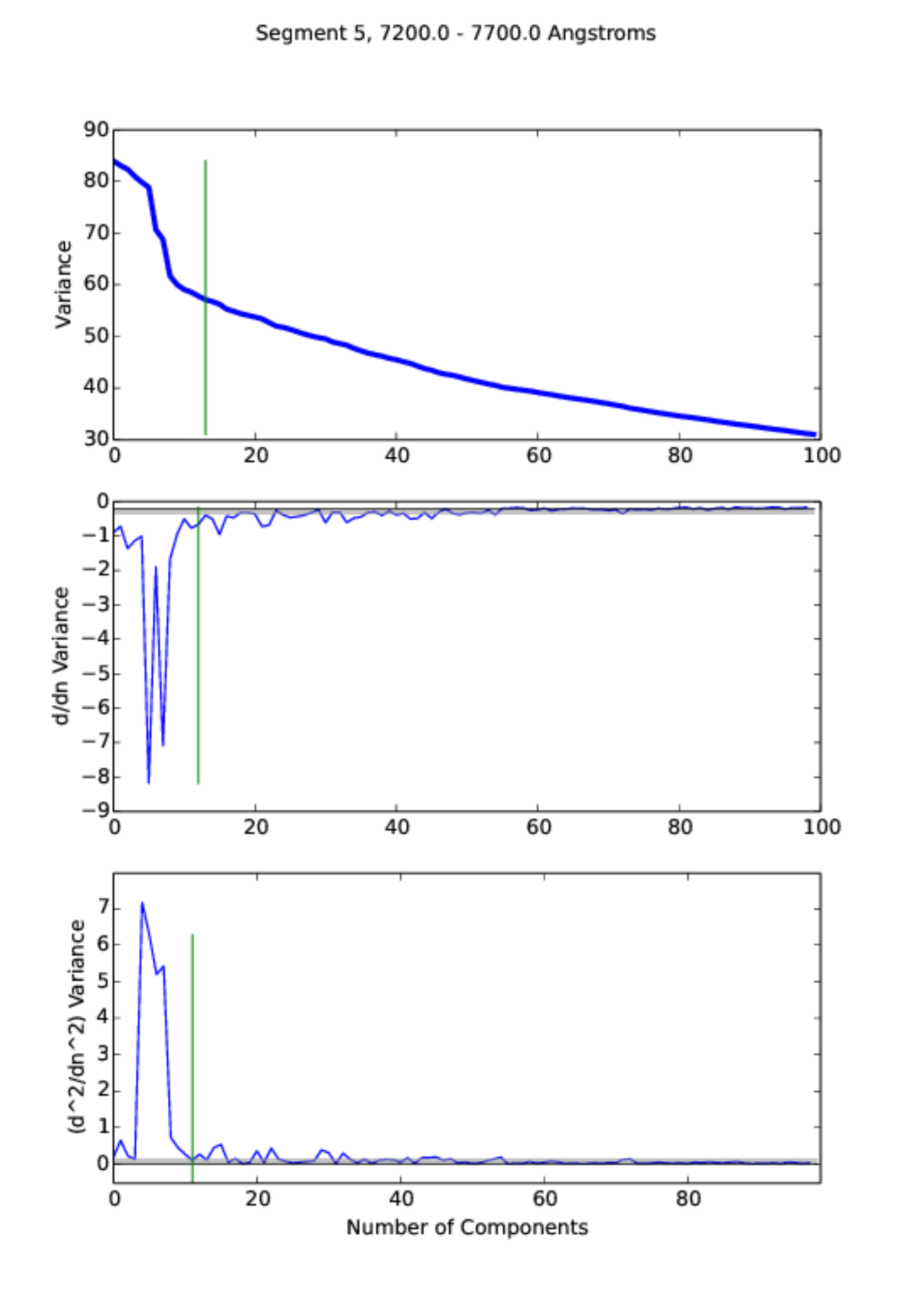}
\caption{\label{fig:varcurve} {\it Above}: The calculated variance curve shows the reduction in the variance with each additional eigenmode. {\it Middle}: The derivative of the variance curve. The initial spikes are associated with the rapid drop in the contribution of the residuals in these eigenspectra. The green line identifies where the curve enters the linear region. {\it Bottom}: The second derivative of the variance curve. This plot identifies the inflection points in the variance curve.}
\end{figure}

This character of the curve divides it into two regions which we use to select the number of eigenspectra, as shown in Figure \ref{fig:varcurve}. The first region (left of the green line) is associated with the most powerful eigenspectra that characterize the sky emission. This part of the eigenspace can also be seen in Figure \ref{fig:evecmatrix} (above the red line), where the eigenspectra are dominated by discrete sky features, for example, that at 7930 \AA.

The second region in Figure \ref{fig:varcurve} (right of the green line) is associated with the remaining eigenspectra that remove remaining source signal and noise. This part of the eigenspace appears also in Figure \ref{fig:evecmatrix} (below the red line), where the eigenspectra take on a more Fourier-transform-like characteristic with eigenvectors that are sinusoidal with increasing frequency in each eigenspectrum.
In this regime the successive eigenvectors begin to remove source signal and noise linearly, which makes the variance curve flatten out, finally ending at zero with the full number of eigenspectra.

The optimization algorithm is tuned to select the number of eigenspectra that represents the point just as the variance curve enters the ``linear" regime. We identify this point by calculating the second derivative of the variance curve and identifying where it approaches zero, which would be the second derivative of a straight line. In a given segment, this transition typically occurs after approximately 10 eigenspectra, as in Figure \ref{fig:varcurve}.
Though the choice of eigenvectors is done autonomously, the user naturally has the choice to include more or fewer eigenspectra.

\begin{figure}
\centering
\includegraphics[width=\linewidth, keepaspectratio=true]{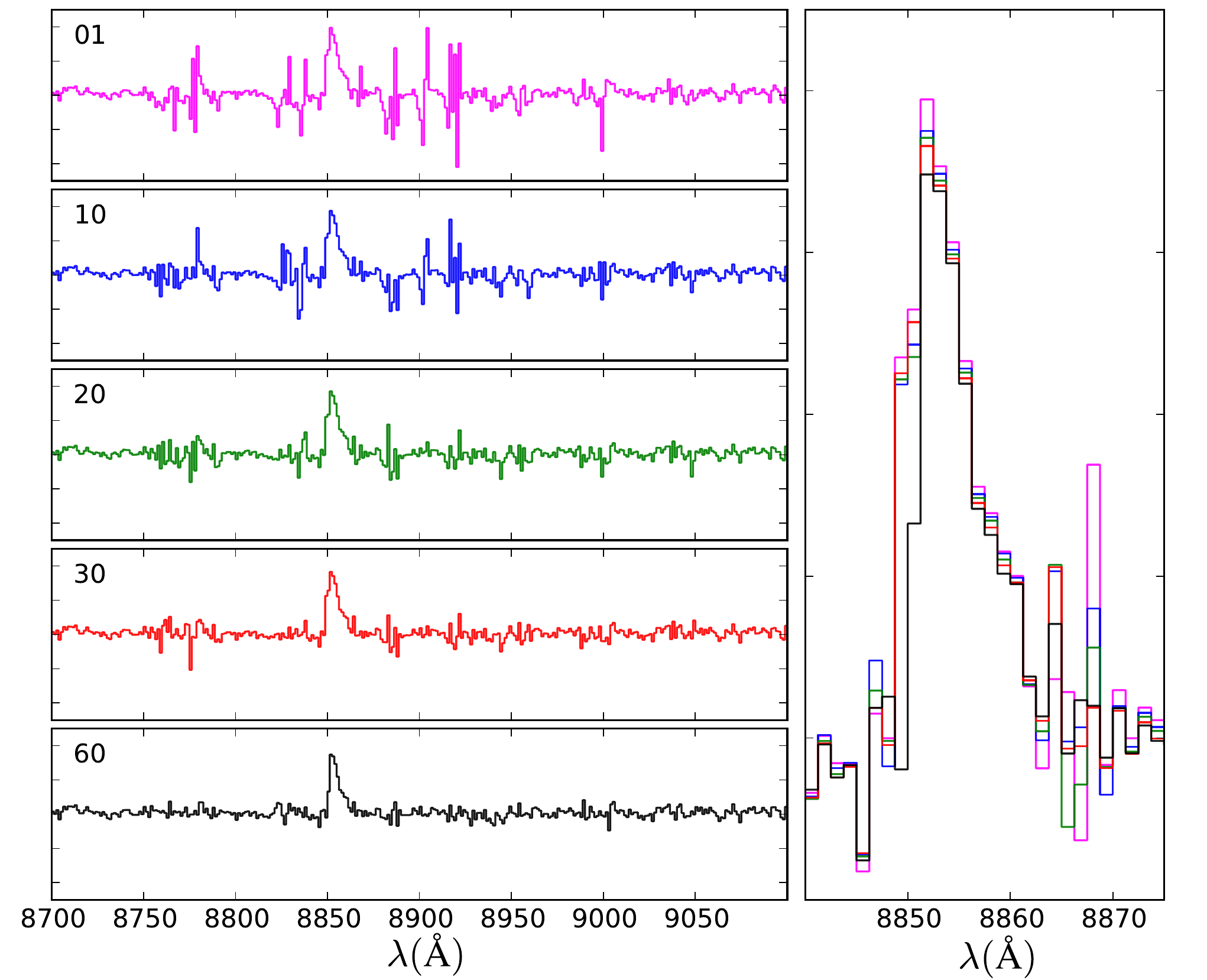}
\caption{\label{fig:varcurvesource} Example of ZAP impact on a galaxy spectrum with Ly$\alpha$ emission line taken from the HDFS MUSE exposure with an increasing number of eigenspectra used (1, 10, 20, 30, 60). As more eigenspectra are used, the eigenspectra become less associated with the sky emission lines and are more likely to affect the source.}
\end{figure}

In Figure \ref{fig:varcurvesource} we illustrate the impact of choosing an increasing number of eigenspectra to characterize the residuals. 
In the left panels, we illustrate the resulting spectra after applying 1, 10, 20, 30, and 60 eigenspectra on the spectrum of an emission line galaxy in the Hubble deep field south \citep{bacon2015HDFS}. 
On the right we show all of the corrected data together. The residuals decrease with an increasing number of eigenspectra, however, once the number of eigenspectra gets too large, the flux and line profile begins to be affected. 

\section{General Use Cases}
\label{sect:cases}

ZAP has been used to improve data in a variety of science investigations, but some consideration must be made when applying it in different situations.
Most science cases can be described by two situations, the ``sparse-field'' case and the ``filled-field'' case. In these two situations the use of ZAP requires a slightly different observing strategy. In the following sections we describe these primary use cases and describe how ZAP can be applied.

	\subsection{Sparse Field case}
	\label{sect:sparse}
	
\begin{figure*}
\centering
\includegraphics[width=0.9\linewidth, keepaspectratio=true]{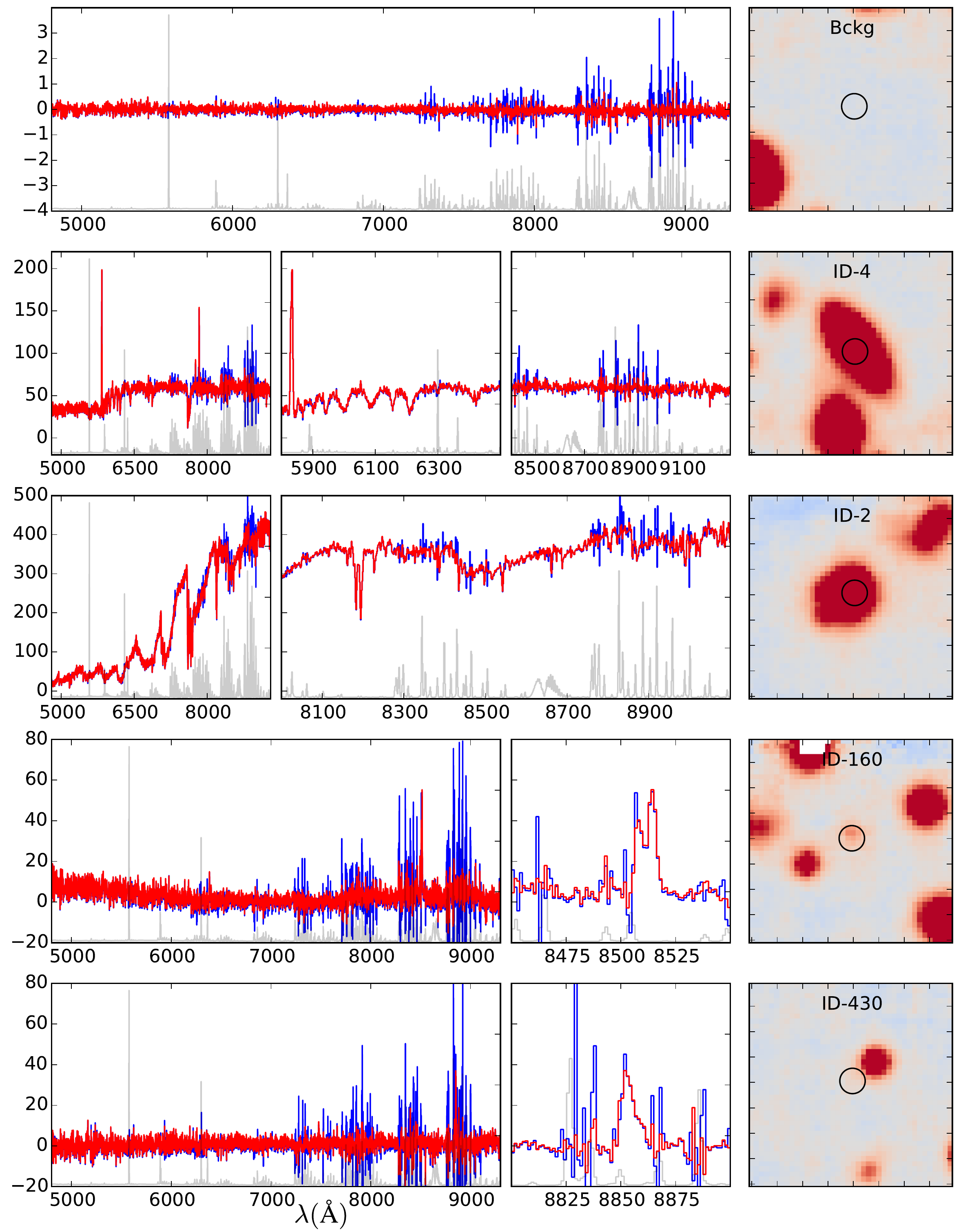}
\caption{\label{fig:sparse} A variety of sources from the HDFS combined datacube. The source numbers refer to the MUSE HDFS catalog \citep{bacon2015HDFS}. The exposure shows that the sky subtraction has not affected the object spectra. All spectra are taken from a 1 arcsec diameter aperture as shown in the right hand images. The blue spectra represent the spectra using the pipeline sky subtraction, the red spectra represents spectra processed with ZAP, and the gray spectra represent the spectrum of the sky. 
{\it Top row}: An empty sky region showing the reduction in residuals throughout the optical range. 
{\it Second row}: A galaxy spectrum with an elaborate continuum shape consisting of many broad absorption and emission lines which are not affected by the process. 
{\it Third row}: A star spectrum with an elaborate shape at wavelengths longer than 7000 \AA.
{\it Fourth row}: An emission line source close to a cluster of sky features at longer wavelengths.
{\it Fifth row}: A weak emission source in the midst of a cluster of sky emission features.}
\end{figure*}
	
In the sparse field case, the field of view contains only faint sources or objects that cover a small angular size. In this case most (e.g. 90\% or more) of the field is dominated by sky signal with only weak sources that are dwarfed by the sky emission lines. In the sparse field case, testing suggests that the spatial masking portion of the ZAP algorithm can usually be safely omitted, but the application of spatial masks will always make the process more robust, allowing a more aggressive reduction in residuals.

In Figure \ref{fig:sparse}, we compare the results of applying ZAP (red spectra) vs the original MUSE pipeline (blue spectra) to sparse field data on a deep observation of the Hubble Deep Field South (HDFS). This dataset is full described in \cite{bacon2015HDFS}. In this test ZAP was applied to the data cube from each individual 30 minute exposure datacube and then again on the final datacube obtained by stacking 53 exposures. The selected ZAP parameters were a filter width of 100 pixels for the generation of the eigenspectra, then a refiltering width of 50 pixels for the calculation of the eigenvalues. The top row shows an empty sky spectrum. The sky residuals are reduced at all wavelengths from 4750 \AA\ to 9350 \AA. 

The remaining panels show representative objects. In the second row of the figure we show the effect on a bright source in the field. In this case, we show that the standard pipeline leaves behind residuals in the red part of the spectrum at wavelengths greater than 7000 \AA; however, with ZAP the source spectrum is not affected and more residuals are removed. 

In the third row of Figure \ref{fig:sparse} we show the effect on a  late-type star in the field. This source is bright relative to the sky residuals, but has an elaborate spectrum, due to the large number of absorption features in the stellar atmosphere.  At red wavelengths the structure of this spectrum is made clearer, making features such as the break at 8800 \AA\ more evident. 

In the fourth row of Figure \ref{fig:sparse} we investigate the spectrum of a weaker emission source with a faint continuum. The continuum shape in the blue is not affected, and the doublet emission feature in the red at 8520 \AA\ is preserved. The narrow sky residuals adjacent to the emission lines and on top of the blue line of the doublet are reduced without affecting the line profiles.

In the fifth row of Figure \ref{fig:sparse}, we investigate a faint emission line object that is in the middle of a sky emission line band. This emission source is flanked by emission line residuals that would otherwise confuse the detection of such a source. When we apply ZAP, the emission line becomes more apparent and the line profile is preserved.
		
	\subsection{Filled Field case}
	\label{sect:filledfield}

ZAP has been developed to also operate for the case where the astronomical source fills the field of view e.g. a nearby galaxy. We find that ZAP works well in this case if a short exposure of a nearby sparse field is used to define both the initial ``zeroth order sky'' (Sect. \ref{sect:zorder}) and also the eigenspectra (Sect. \ref{sect:eigenvectors}). The eigenvalues are then calculated in the usual way from the filtered science frame (Section \ref{sect:eigenvalues}).
The observation strategy for this case therefore requires measurement of the sky in nearby blank fields that are mostly absent of sources. 
By using these external sky frames, ZAP excludes the source spectrum entirely from the eigenspectra and zeroth order sky subtraction. 

The observations should be performed in an object-sky-object pattern so that the sky is sampled close in time. 
This approach simulates the nodding strategy used for near infrared instruments, but allows the observer to take shorter exposures of the sky compared to the target exposures. 
The degree to which high level sky subtraction is needed will depend on the particular science case, but we have been successful with 900 s exposures on source and 120 s exposures on the adjacent sky. This particular strategy therefore amounts to only a 6\% increase in total exposure time, compared to a factor of 4 increase when using a conventional pixel by pixel chopping scheme to obtain a similar signal to noise ratio. This is because the eigenvector-based sky subtraction is effectively noise free, while the sky exposures in a pixel by pixel subtraction will introduce its own noise.

To apply ZAP on a filled-field source, we simulate the sparse field situation through the application of continuum filtering to the data cube. After subtracting the initial zero level sky spectrum constructed from the offset frame, we then apply the weighted median continuum filtering algorithms with a small 20-50 pixel window to eliminate most of the source signal. When this filter is applied, the remaining signal is in only a few narrow astronomical features such as emission or absorption lines and residual continuum break derivatives, plus the remaining sky residuals. The data are then projected onto the eigenspace via a dot product of the eigenspectrum matrix with the filtered data to calculate the eigenvalues. 

\begin{figure}
\centering
\includegraphics[width=0.9\linewidth, keepaspectratio=true]{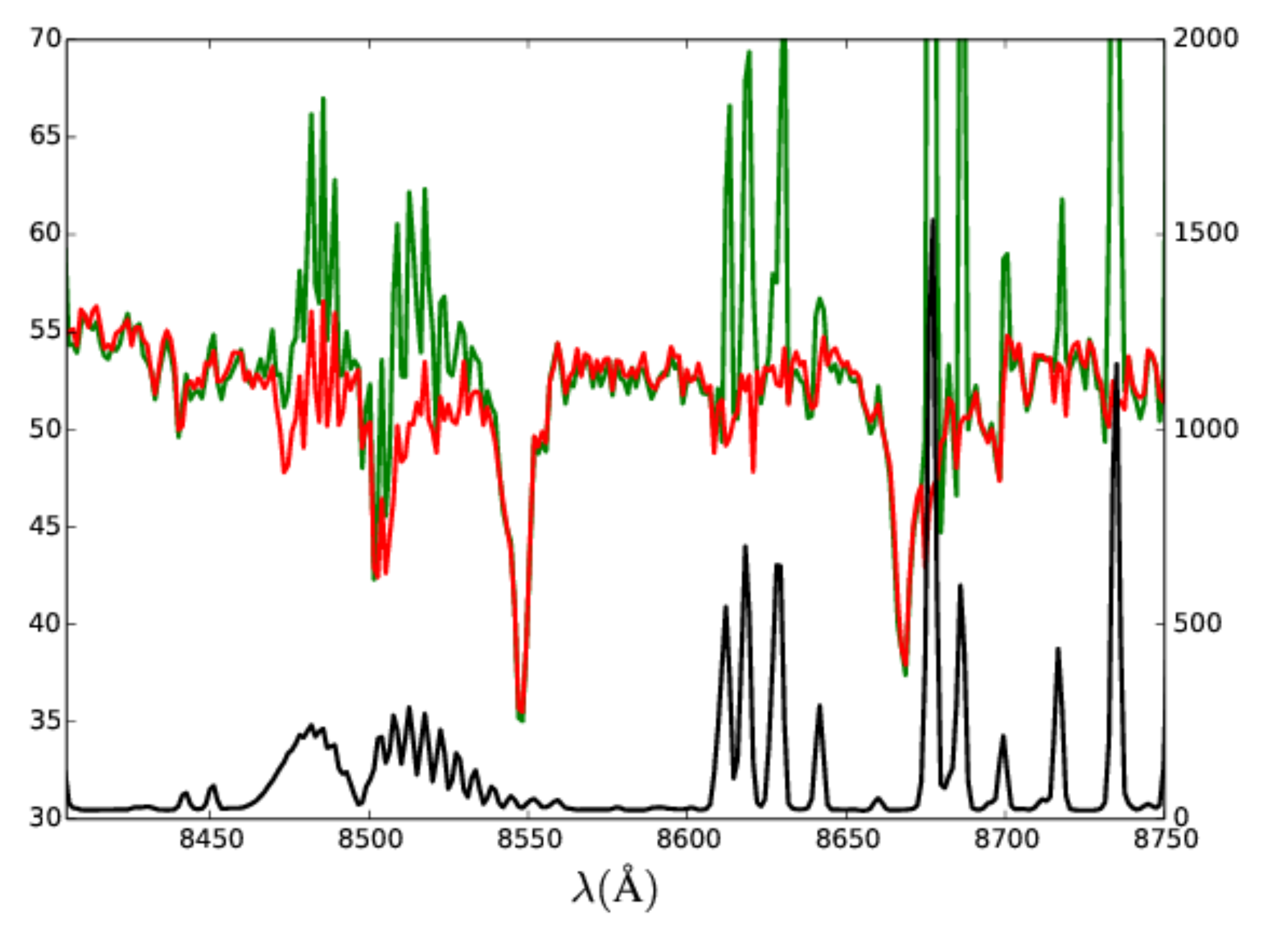}
\includegraphics[width=0.9\linewidth, keepaspectratio=true]{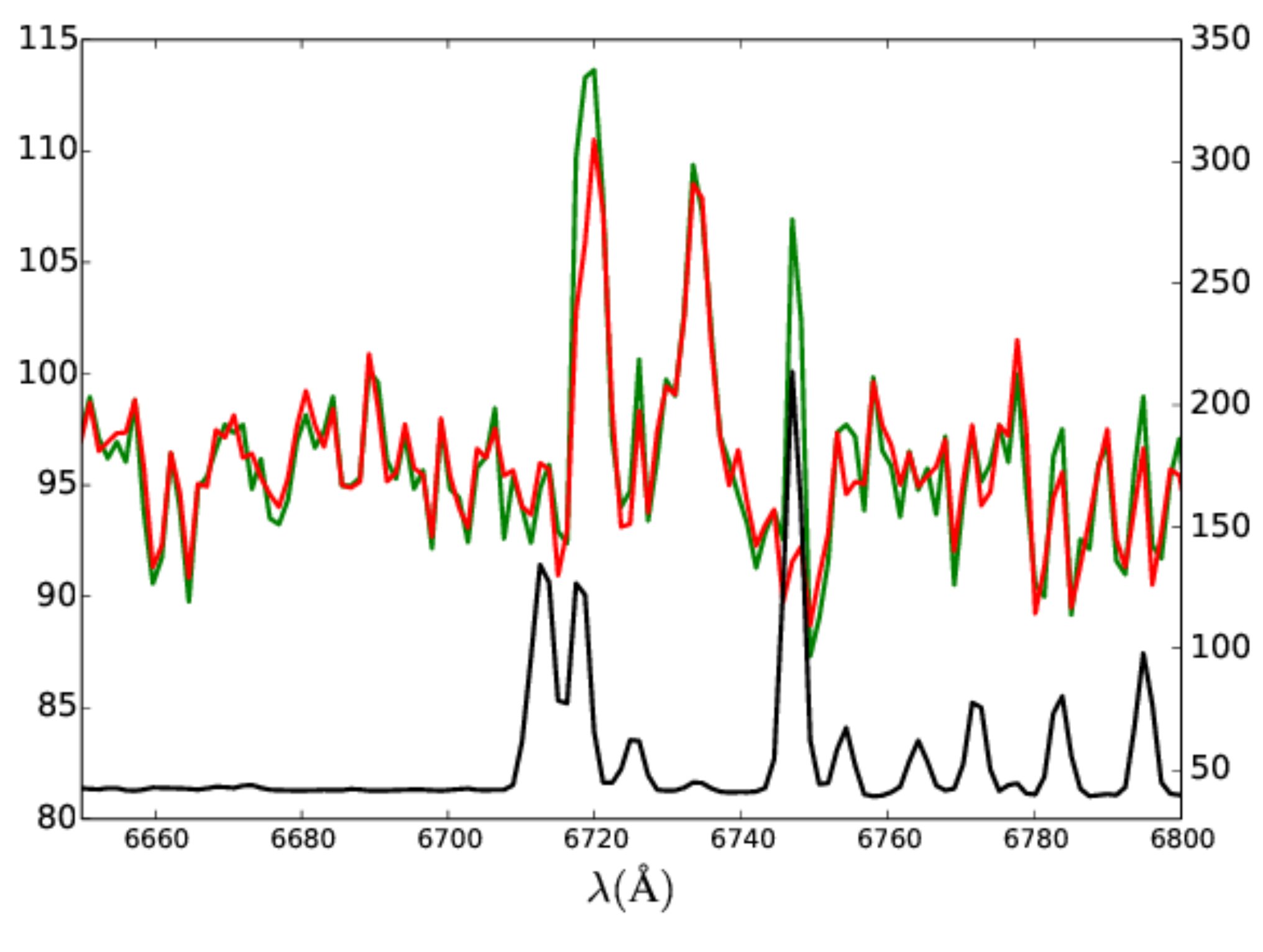}
\caption{\label{fig:ulirg} {\it Above}: The green spectrum shows the result of performing a single spectrum removal of the sky spectrum, similar to the dedicated fibre approach to sky subtraction in a region associated with the Calcium triplet. The red spectra show the same set of spaxels using ZAP as a complete sky subtraction method.  The black spectrum shows the spectrum of the sky emission in the same location and part of the spectrum. The fluxes are all in $10^{-20}~$erg s$^{-1}$cm$^{-2}$, with the scale for the remaining sky residuals on the left vertical axis and the scale for the original sky spectrum (plotted in black) on the right vertical axis. {\it Below}: The same legend as the above plot.  These emission lines are in the blue end of the spectrum where the density of sky lines is lower. The residuals are still removed.}
\end{figure}

In Figure \ref{fig:ulirg} we show the result of applying ZAP to an exposure of a nearby ultra luminous infrared galaxy using this method. In each of these plots, the green lines shows the result of subtracting the average sky from an adjacent sky frame. The red lines show the result of ZAP on the same system. In each case, the residual sky features have been substantially reduced resulting in a cleaner spectrum even at longer wavelengths.

\section{Parameter selection}
\label{sect:quality}
	
There are very few parameters to adjust when applying ZAP, and only a few of these can strongly affect the result.
The two most important are the choice of spatial masks to exclude known objects from the eigenvector construction, and the \texttt{cfwidth} parameter that controls the width of the filter used for the further sanitization of the eigenspectra (Section \ref{sect:eigenvectors}) and/or the calculation of the eigenvalues (Section \ref{sect:eigenvalues}).  
In both cases the goal is to keep the eigenspectra and reconstructed sky residual spectrum free of signal from the astronomical sources.	
	
	\subsection{Sky measurement}
	\label{sect:sky measurement}

In any sky subtraction algorithm, the only way to subtract the sky signal reasonably is to have some measure of the sky signal. 
For ZAP, the sky signal can come from the image itself (sparse-field case) or from an exposure taken close in time (filled-field case), with each case requiring a different consideration of observing strategy.

In the sparse field case, the signal of the sources are often dominated by the signal of the sky. 
As the density and flux of the sources increase, masking of the sources becomes more relevant to the process. 
Best practices are to run ZAP iteratively and to apply masks to sources that are successively detected as the sky residuals are removed. ZAP will work best when the number of spaxels included in the zeroth order sky spectrum and eigenspectrum calculation is at least a factor of 10 greater than the number of pixels in the spectral axis for the largest spectral segment. For a single 90,000 spaxel MUSE exposure, this corresponds (with ten spectral segments) to needing only $\sim10\%$ of the field of view for eigenspectrum calculation, i.e. 90\% of the field of view could in principle be masked.  This is another way to appreciate the importance of segmenting the wavelength range, since without this, we would need most of the spaxels to be usable to achieve this factor of ten margin.  However, better results are obtained with larger regions of sky because more measurements are used to find the correlations in the residuals and define the zeroth order spectrum. In the minimal case of using only a small fraction of the field of view, the sky region should be if possible distributed over the field of view so as to well sample the different optical paths in the instrument. 

In the filled field case, the problem becomes temporal, due to the variability of the sky emission. If the sky exposure is taken at a time too distant from the object exposure, the zero level sky will be too different from the zero level in the object exposure due to variations in the sky signal with time. While the eigenspectra will correct some of the variations, the variations will be more distant from the mean signal and make residuals more difficult to characterize. This ultimately will correspond to larger residuals.

\subsection{Continuum filter width (\texttt{cfwidth})}
\label{sect:cfwidth}

The \texttt{cfwidth} parameter determines the size of the filtering box used for the weighted median filter. The filter is designed to remove source signal and ignore any remaining sky emission residuals in the spectrum. This filtering is not perfect due to issues such as source signal to noise ratio and interference in regions where sky residual features are closely packed. In general, the width of this filter can be set by the densest cluster of sky line residuals in the spectrum. For MUSE this happens between the wavelengths of 8450 \AA\ to 8550 \AA. This implies that a filter width of 100 pixels is sufficient in most cases. Setting the filter width too small at this stage risks the removal of sky features along with the source, therefore removing them from the eigenspectra calculation.

Through extensive empirical investigations into the results of applying ZAP, we found that the continuum filter width can be set even smaller in the eigenvalue calculation step. With a cfwidth of 20-50 pixels, we are able to reject sources and therefore avoid influence on the sources when the data is projected onto the eigenvalues.
	
\section{Discussion}
\label{sect:discussion}

In extensive testing and application to MUSE datacubes, we have found that ZAP works very well. In this section we provide a short discussion of why we think this and also mention a few known problem areas.

ZAP works well with respect to the standard pipeline because it takes advantage of the large number of more or less independent spectra to characterize the residuals.
The number of spectra (90,000) in a single exposure is equivalent to some entire spectroscopic surveys. 
With this large volume of spectra, the correlations between residuals are more able to stand out in the calculated eigenspectra. 

The full set of 3,680 eigenspectra (plus the associated eigenvalues) completely characterizes all of the unfiltered information in the datacube. However, typically only 2-3\% of the eigenspectra (in each segment) are used to construct the sky residual spectrum. Put another way, 2-3\% total ``information'' in the cube is used to characterize 25\% of the variance, which is empirically dominated by the sky residuals. The measure we use to determine the number of eigenspectra used is connected to the change in average variance of the entire datacube, thereby using the entire field to constrain the number used.

One of the strengths of ZAP is that the reconstruction of sky residuals is completely independent of spatial position. 
ZAP makes no assumption of any spatial coherence in either the identification or correction of sky residual features and makes no use of any such coherence in constructing the sky residual cube.
This is unlike the case in e.g. long-slit spectroscopy, where it is usually assumed that the sky spectrum varies continuously along the slit. This has proved especially important for MUSE because the complex optical path introduces sharp spatial discontinuities in the sky residuals.  This additionally means that the masking of faint objects is less critical as any remaining astronomical signal from them will be diluted over the whole field rather than just a small spatially coherent area such as a long slit. Furthermore, by using only the strongest few eigenspectra, these weak and highly diluted astronomical signals will likely not enter at all into the sky residual spectra.

Since ZAP does not depend explicitly on a model of the sky spectrum, but rather on the correlated residuals, it can be run more than once.  A successful application of this has been to apply ZAP to each exposure of an observation, and then to apply ZAP again to the co-added datacube. This specific application is useful in the ``sparse-field'' case where the zero level sky spectrum and the eigenspectra are calculated from the same data cube as the sources. In the final application the masks become more important, since the sources can become major contributors to the overall variance, but improved sky subtraction means those masks can be better defined for the sources. For example, if an extended emission region becomes evident, this region can be masked to ensure it is not included in the eigenspectra. This iterative application differs from simply selecting more eigenspectra because it allows the basis vectors of what were initially weak residuals to be re-characterized, and better isolated into a small number of eigenspectra.

The application of ZAP does have some costs. In some cases if too many eigenspectra are used, ZAP can have an effect on the noise within the datacube itself. 
This de-noising of the datacube will generally occur in the small ranges of wavelengths that are associated with the eigenspectrum peaks rather than over the whole wavelength range.
In other words, there is a tendency to artificially de-noise the datacube at the location of strong emission lines, simply because the eigenspectra will have strong features at these wavelengths to deal with the residuals.
This can be recognized by comparing the noise in the ZAP processed cube with the ideal estimate of the variance (based on Poisson statistics and the drizzling pattern) that is produced by the MUSE pipeline. 

One skyline, [OI]5577, presents an extreme example of this type of de-noising. This line is completely isolated and the residuals associated with it are therefore uncorrelated with those of other lines, since it completely dominates its wavelength segment (Fig. \ref{fig:skyspec}) and does not have an association with adjacent sky emission lines. In this case, the eigenspectra that characterize the residual sky spectrum are not constrained by the projection onto emission lines at other wavelengths within a spectral segment. In this case, the eigenspectra will be able to aggressively remove all traces of the sky emission line including Poisson noise and any previously unfiltered astronomical signal. This means that caution should be taken when using empirical measures of noise from the datacube itself. Fortunately, this emission line is unusual in that the other sky emission lines have associated sky emission and correlated residuals within their respective spectral segments (Fig. \ref{fig:skyspec}). 

A feature of ZAP that may have struck readers as strange, and which indeed can pose a problem, is the refiltering with different sized windows in the eigenspectrum calculation and eigenvalue calculation steps. Two sizes are used because of the different goals of those particular steps in the process.  The first filtering step is aimed at isolating the sky residuals which can be identified over the whole datacube. In this case the larger filter box is needed in order to traverse regions of densely packed residuals. Having characterized most of the continuum, the correlation between the residual characteristics is determined using the abundance of spaxels in the datacube. A quick check of whether or not the filter width and masking has been sufficient in the construction of the eigenspectra is to investigate the first few eigenspectra and determine if there are significant low frequency components. If these features are present and have a significant strength compared to the narrow residual features then this suggests, a better mask or a filter with a smaller width should be applied.

The second filtering step is aimed at calculating the eigenvalues for every observed spectrum. 
In this case, residual astronomical signals that have not been filtered out can affect the calculation of an eigenvalue if the signal is coincident with the features in an eigenspectrum. 
A smaller weighted median filter will trace and subtract the astronomical signal better but still leave behind narrow sky residual features.
These narrow features will dominate the magnitude of the calculated eigenvalues leading to a better residual sky subtraction.
A notable exception is the O$_2$ 1-0 sky emission from 8610 \AA\ to 8706 \AA\ where the density of sky emission lines forms a continuous broad feature. In this case, a small 20 pixel window filter has difficulty distinguishing sky and source signal in the spectrum, so it will remain in the final data product.
	
\section{Conclusion}
\label{sect:conclusion}

In this work we have developed a PCA based method for the purpose of sky subtraction in integral field spectroscopy. Developing ideas designed for multi-fibre survey spectroscopy, we have developed ZAP, a general use tool which is able to substantially reduce sky subtraction residuals in complex integral field spectrographs such as MUSE.  

The refinements are centered on keeping the calculated eigenspectra from influencing the eigenspectra.
Inclusion of weighted median filtering and spatial source masking help in combination to sterilize the eigenspectra from astronomical signals. In the case of MUSE, the huge number of spectra also helps to isolate the sky residual features into the first few eigenspectra.

The segmentation of the dataset into discrete wavelength regions allows the SVD to operate on the variations of the sky emission species independently and has the added benefit of substantially speeding up the code by reducing the size of the matrix inversions. The inclusion of an autonomous eigenvalue selection algorithm further helps to select a number of modes that will minimize the impact on astronomical signal and the noise. 

Furthermore, the generality of ZAP allows it to be used in many observational situations including sparse fields with faint sources as well as filled fields dominated strong sources. In the latter case, observing efficiency is improved by minimizing the amount of time spent observing adjacent blank fields.

Together, the improvements introduced by and included in ZAP allows the technique to be used in almost all observational situations to reduce sky residuals, and therefore help many scientific studies.   ZAP can be used as a stand alone post-processing sky subtraction method or as a supplement to clean up the residuals produced in previously sky subtracted datacubes. 

While some features of MUSE are particularly well suited to a PCA approach, ZAP is likely to be of general use in the processing of all types of datacubes or spectroscopic surveys affected by similar sky residuals. This software can be found on the Astrophysics Source Code Library \citep{zapASCL} and is available for download and use at \mbox{\texttt{http://muse-vlt.eu/science/tools}}.

\section*{Acknowledgements}
This research is supported by the Swiss National Science Foundation. KS acknowledges Anat Stolarsky, Maryam Shirazi, and Aseem Paranjape for fruitful discussions during development. RB and SC acknowledge support from the ERC advanced grant 339659-MUSICOS.



\bibliographystyle{mnras}
\bibliography{zap} 








\bsp	
\label{lastpage}
\end{document}